\documentclass[a4paper,11pt]{article}
\usepackage{pos}

\usepackage{lineno}

\AtBeginShipoutNext{\AtBeginShipout{%
    \ifnum\thepage=7\AtBeginShipoutDiscard\fi}}

\title{Dielectron measurements in Au+Au collisions at BES-II energies with the STAR experiment}

\author*[1]{\textit{Chenliang} Jin}

\affiliation[]{for the STAR Collaboration}
\affiliation[1]{Department of Physics and Astronomy, Rice University,\\
  6100 Main St, Houston, TX, 77005, USA}

\emailAdd{cj17@rice.edu}

\abstract{Dielectrons, emitted during the evolution of the hot and dense QCD medium created in relativistic heavy-ion collisions, offer an effective probe of the hot medium properties, as they do not involve strong interactions. The dielectron emission rate is proportional to the medium's electromagnetic spectral function. In the dielectron mass range from $400$ to $800$ MeV/$c^{2}$, the spectral function probes the in-medium $\rho$ meson propagator which is sensitive to the medium’s properties including the total baryon density and the temperature. By measuring thermal dielectron production, we can study the microscopic interactions between the electromagnetic current and the medium. The RHIC Beam Energy Scan (BES) program provides a unique opportunity to systematically study dielectron production in a collision energy range where the total baryon density and temperatures are varying substantially. In these proceedings, STAR measurements of thermal electrons produced in Au+Au collisions at $\sqrt{s_{\text{NN}}}=$ 7.7, 9.2, 11.5, 14.6 and 19.6 GeV will be reported. The results will include the thermal dielectron spectra, differential/total excess yield, and the temperature extracted from the low invariant mass range, as well as their collision energy dependence.}

\FullConference{%
  HardProbes2024\\
  22-27 September 2024\\
  Nagasaki, Japan
}

\begin{document}
\maketitle

\section{Introduction}

Hot and dense strongly interacting matter including Quark-Gluon Plasma (QGP) and hadronic resonance gas phase can be produced in the laboratory by colliding nuclei at high energies. The thermal radiation emitted throughout all stages of the evolution of the system allows us to investigate various properties of the QCD medium. Experimentally, this thermal radiation can be detected via dilepton production, which can reveal the interaction mechanism between the electromagnetic (EM) current and the medium.

In a strongly interacting thermal equilibrium medium, the dielectron emission rate in four dimensional space and momentum  \cite{Feinberg:1976ua, McLerran:1984ay} can be written as Eq. \ref{eq_dilep_rate}.
\begin{equation}
\label{eq_dilep_rate}
\frac{dR_{e^+e^-}}{d^4x \, d^4q} = -\frac{\alpha^2_{\text{EM}}}{3\pi^3 M^2} f_B(q_0, T) g_{\mu \nu} \, \text{Im} \, [\Pi^{\mu \nu}_{\text{EM}}(M_{ee}, q; T, \mu_B)]
\end{equation}
where $f_{B}$ is thermal Bose–Einstein distribution, $M_{ee}$ is the dielectron invariant mass. The emission rate is connected to the imaginary part of correlation function $\text{Im} \, [\Pi^{\mu \nu}_{\text{EM}}(M_{ee}, q; T, \mu_B)]$ defined via the hadronic EM current. In the intermediate mass range ($M_{ll} > 1$ GeV/$c^{2}$), the EM current predominantly exhibits partonic behavior. Conversely, in the low mass range ($M_{ll} < 1$ GeV/$c^{2}$), the current can be rewritten according to the vector meson isospin eigenstate \cite{Sakurai:1969}. Essentially, the EM spectral function is the composition of imaginary components of vector meson propagators, manifesting as $\rho$-dominance \cite{vanHees:2006ng}.

The $\rho$ meson propagator is sensitive to the medium’s properties including the total baryon density and the temperature \cite{Rapp:2000pe}. The RHIC Beam Energy Scan (BES) program provides a unique opportunity to systematically study dielectron production and measure the medium properties in a collision energy range where the total baryon density and temperatures vary substantially \cite{STAR:2017sal}.

\section{Experiment and Analysis}

In these proceedings, preliminary results obtained from Au+Au collisions at $\sqrt{s_{\text{NN}}}=$ 7.7, 9.2, 11.5, 14.6,  and 19.6 GeV with the STAR detector at RHIC are presented. These data are taken in 2019, 2020, and 2021 during Beam Energy Scan Phase-II (BES-II). BES-II provides approximately 10 times the statistics of BES-I, reducing the statistical error by a factor of 4 at $\sqrt{s_{\text{NN}}} = 19.6$ GeV.

The main sub-detectors utilized for particle identification with the STAR framework are the Time Projection Chamber (TPC) and the barrel Time Of Flight (TOF) \cite{STAR:2002eio}. The TPC system measures the ionization energy loss ($dE/dx$)
in the TPC gas with a mean ($dE/dx$) resolution of about 7\% 
and charged particle momenta
with a resolution of $\sim$ 1\% at $p_T$ $\sim$ 1 GeV/c for tracks originating from the collision vertices. 
The TOF system measures the time of flight with a resolution of about 80 ps. It is further combined with the track path length and momentum measured by the TPC to determine the track velocity and then reconstruct the mass to identify the electrons.

The $e^+ e^-$ mass continuum is reconstructed by combining electrons and positrons from the high purity electron candidates. The combinatorial and correlated background are estimated via geometric mean of like-sign pairs with pair sign acceptance correction (PSAC) \cite{PHENIX:2009gyd}. The raw signal is derived by subtracting the like-sign background spectrum, subsequently applying efficiency corrections to yield the fully efficiency-corrected spectrum \cite{STAR:2015tnn}.

Isolation of the dielectron excess requires the subtraction of the known physics background (hadronic cocktail). In this thermal dielectron analysis, the physical background includes direct and Dalitz decays of $\pi^{0}$, $\eta$, $\eta'$, $\omega$, $\phi$, and
$J/\psi$; semi-leptonic decay of correlated charm; and Drell-Yan process.

\section{Dielectron as Spectrometer}

\begin{figure}
\begin{centering}
\includegraphics[width=0.55\textwidth]{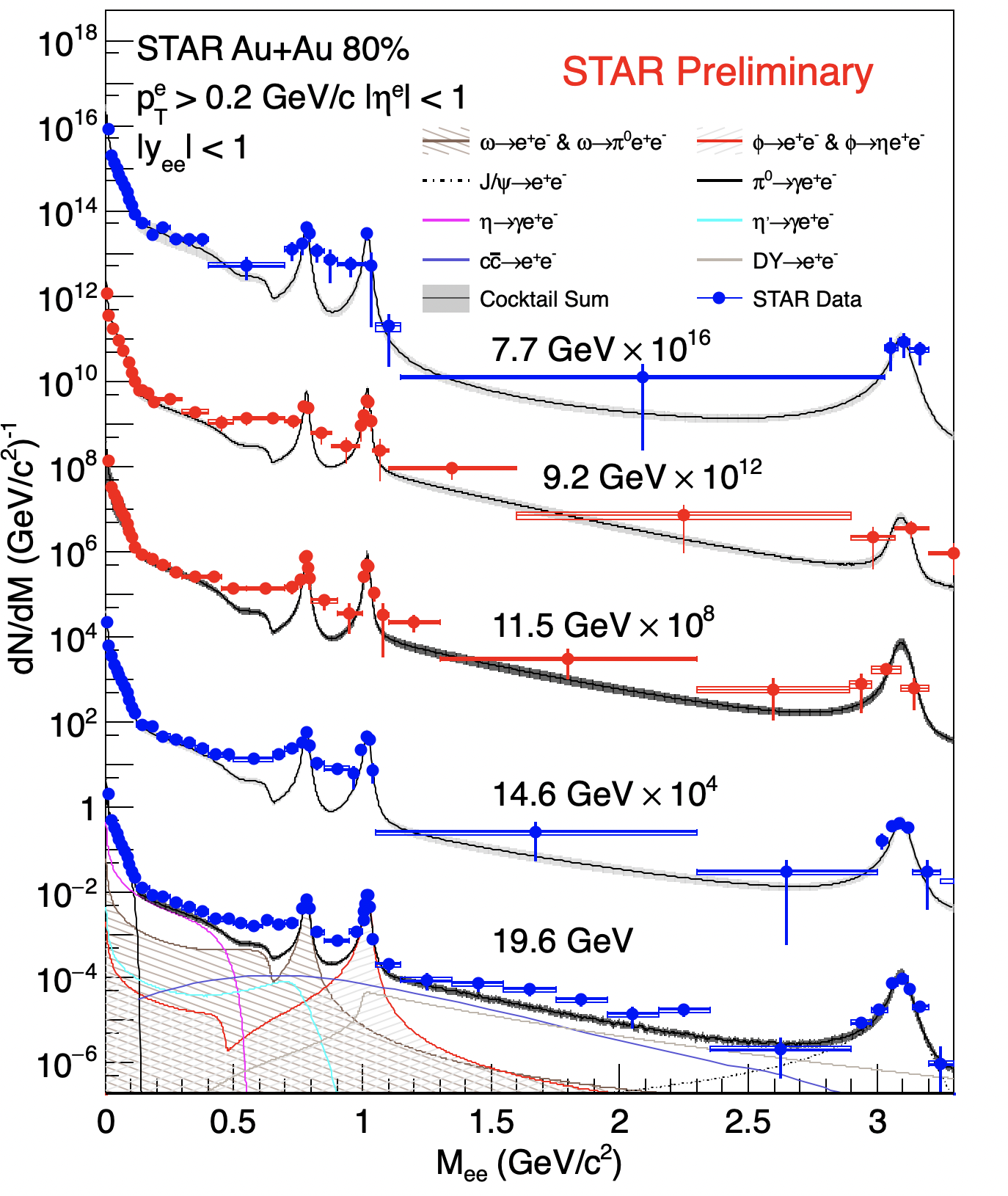}
\par\end{centering}

\protect\caption{Efficiency-corrected dielectron invariant mass spectra within the STAR acceptance for Au+Au collisions at $\sqrt{s_{\text{NN}}}=$ 7.7, 9.2, 11.5, 14.6 and 19.6 GeV. Experimental results with statistical and systematic uncertainties are shown together as points. The physical background is illustrated with solid lines and shadow area refers to hadronic cocktail uncertainty.}

\label{fig_inclusive_spectrum}
\end{figure}

The efficiency-corrected invariant mass spectra for Au+Au minimum-bias (0-80$\%$ centrality) collisions at $\sqrt{s_{\text{NN}}}=$ 7.7, 9.2, 11.5, 14.6, and 19.6 GeV are shown in Fig. \ref{fig_inclusive_spectrum}. The blue and red points represent invariant mass measurements within the STAR detector’s acceptance at mid-rapidity: $|\eta^{e}| <$ 1, $|p^{e}_T| >$ 0.2 GeV/c and $|y^{ee}| <$ 1. The experimental data are displayed together with the calculated physical background, indicated by a solid black line.

\begin{figure}
\begin{centering}
\includegraphics[width=0.48\textwidth]{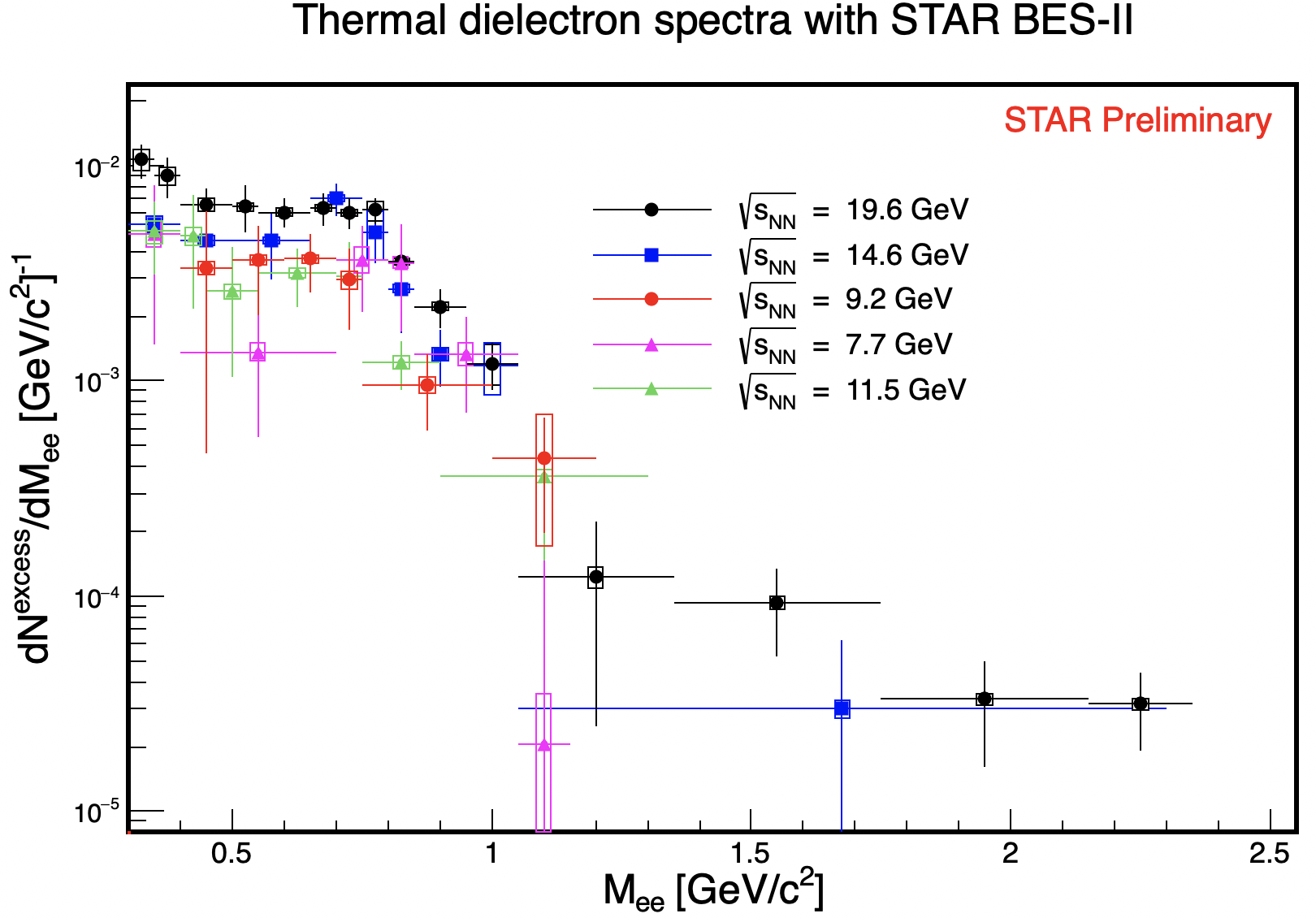}
\includegraphics[width=0.48\textwidth]{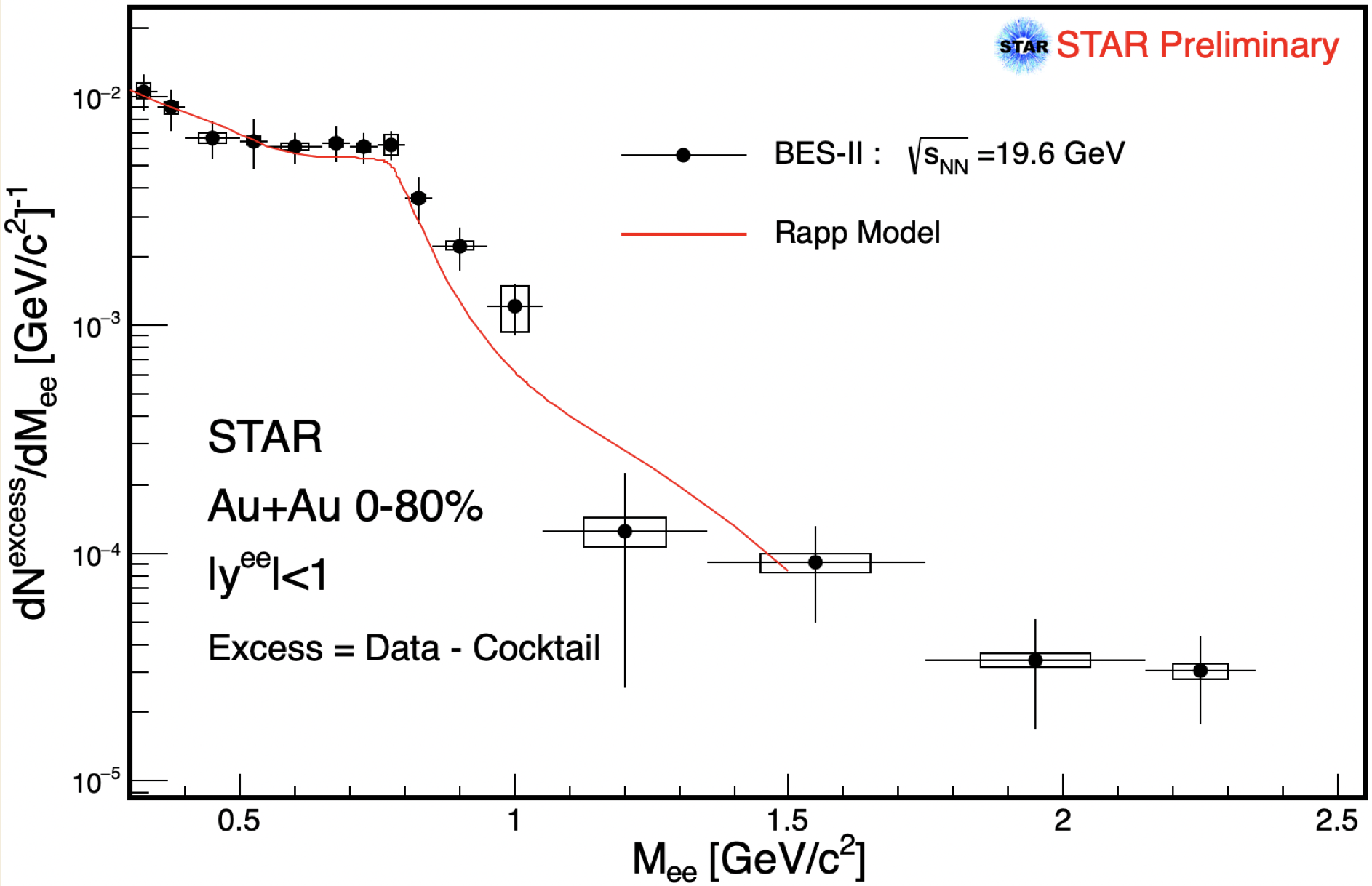}
\par\end{centering}

\protect\caption{Left: Acceptance-corrected excess yield invariant mass spectra (points) for Au+Au collisions at $\sqrt{s_{\text{NN}}}=$ 7.7, 9.2, 11.5, 14.6 and 19.6 GeV. Right: Comparison with the Rapp model calculation (red solid line) at $\sqrt{s_{\text{NN}}}=$ 19.6 GeV.}

\label{fig_exclusive_spectrum}
\end{figure}

Acceptance corrections are crucial for comparing the thermal radiation spectra with other experimental results and minimizing biases from the STAR detector environment after subtracting the physical background \cite{STAR:2015zal}. The left panel of Fig. \ref{fig_exclusive_spectrum} displays acceptance-corrected excess yield spectra at BES-II energies. The right panel of Fig. \ref{fig_exclusive_spectrum} depicts a comparison between the excess yield and Rapp's model prediction at 19.6 GeV \cite{vanHees:2006ng}, underscoring the effectiveness of the many-body theory in capturing the dynamics of the medium.

\begin{figure}
\begin{centering}
\includegraphics[width=0.53\textwidth]{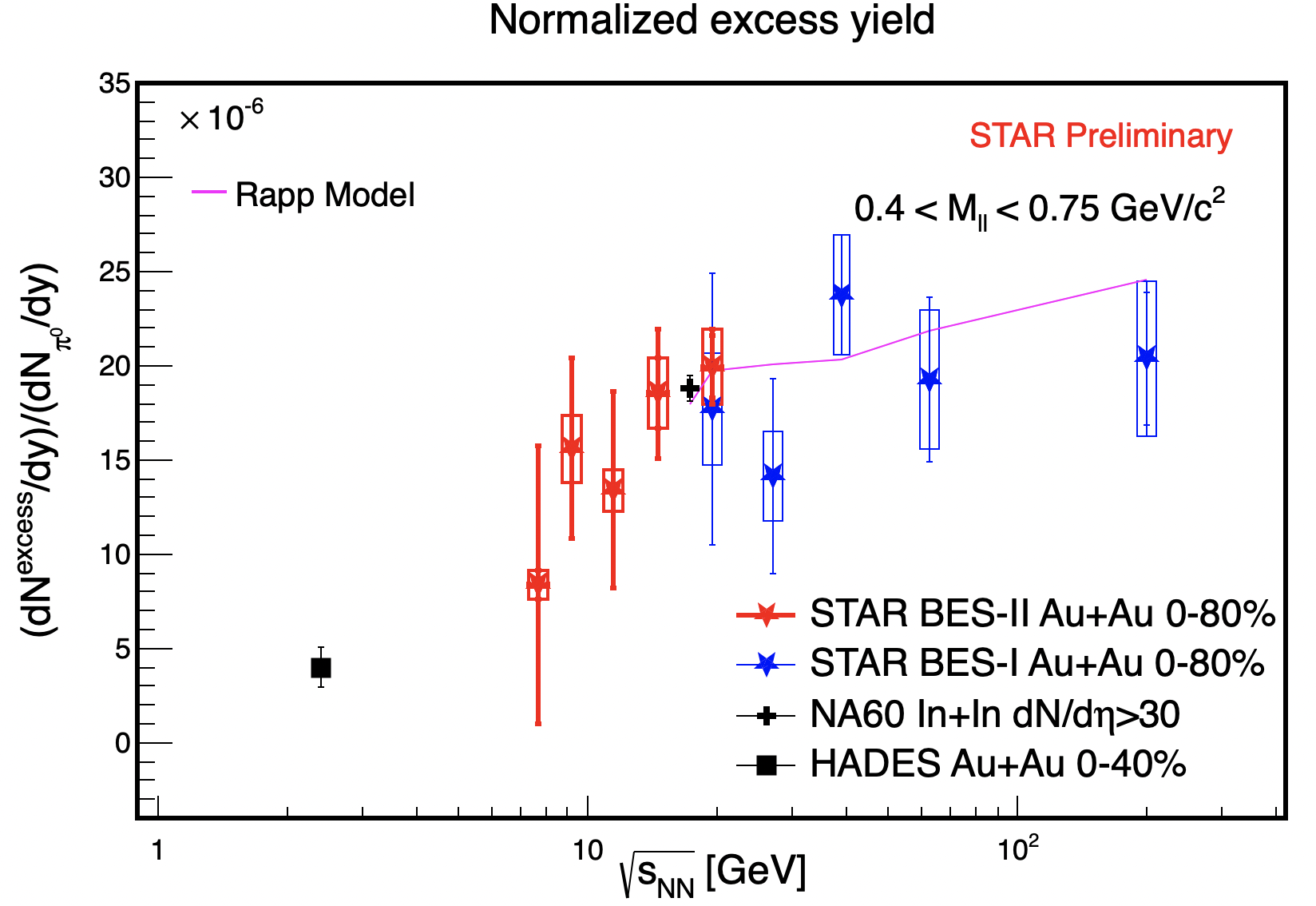}
\par\end{centering}

\protect\caption{\textbf{Integrated excess yield vs. collision energy.} $\pi^{0}$ yield normalized integrated excess yield from different experiments (STAR BES-I: blue star \cite{STAR:2015zal, STAR:2023wta}; STAR BES-II: red star; NA60: black cross \cite{NA60:2008dcb}; HADES: black square) \cite{HADES:2019auv} for 0.4 $< M_{ll} <$ 0.75 GeV/$c^2$ with Rapp model prediction (pink solid line) \cite{Rapp:2000pe}. }

\label{fig_excess_snn}
\end{figure}

The $\pi^{0}$ yield (estimated by $N_{\pi^{0}} = (N_{\pi^{+}}+N_{\pi^{-}})/2$) normalized integrated excess yield for 0.4 $< M_{ll} <$ 0.75 GeV/$c^2$ as a function of the collision energy is presented in Fig. \ref{fig_excess_snn}.
New results from STAR BES-II suggest a downward trend in the normalized integrated excess yield as collision energy decreases. It is contrary to initial expectations \cite{STAR:2014note} that 
normalized integrated excess yield will increase with higher total baryon density, which could offer constraints for theoretical models.

\section{Dielectron as Thermometer}

\begin{figure}
\begin{centering}
\includegraphics[width=0.48\textwidth]{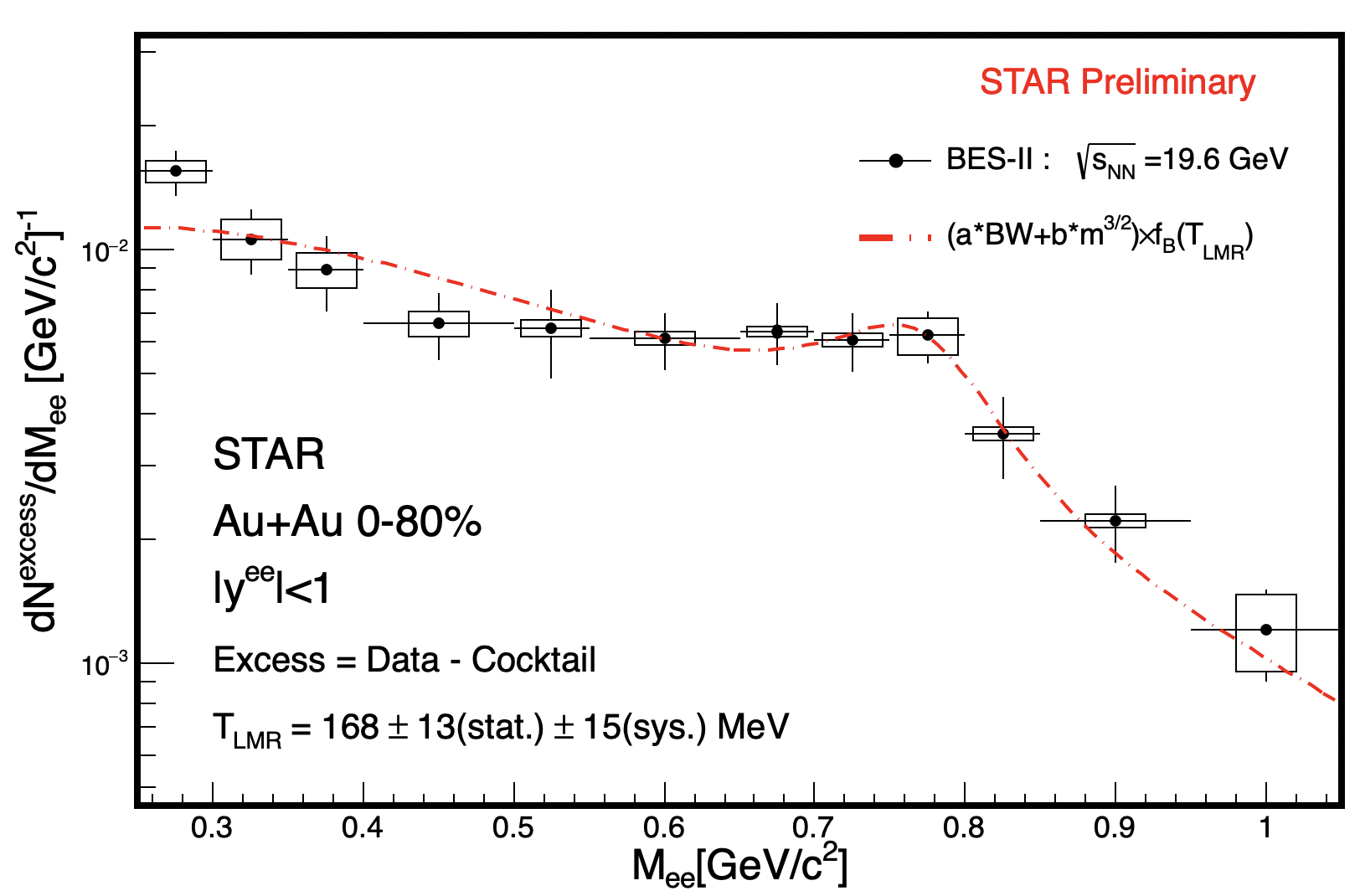}
\includegraphics[width=0.48\textwidth]{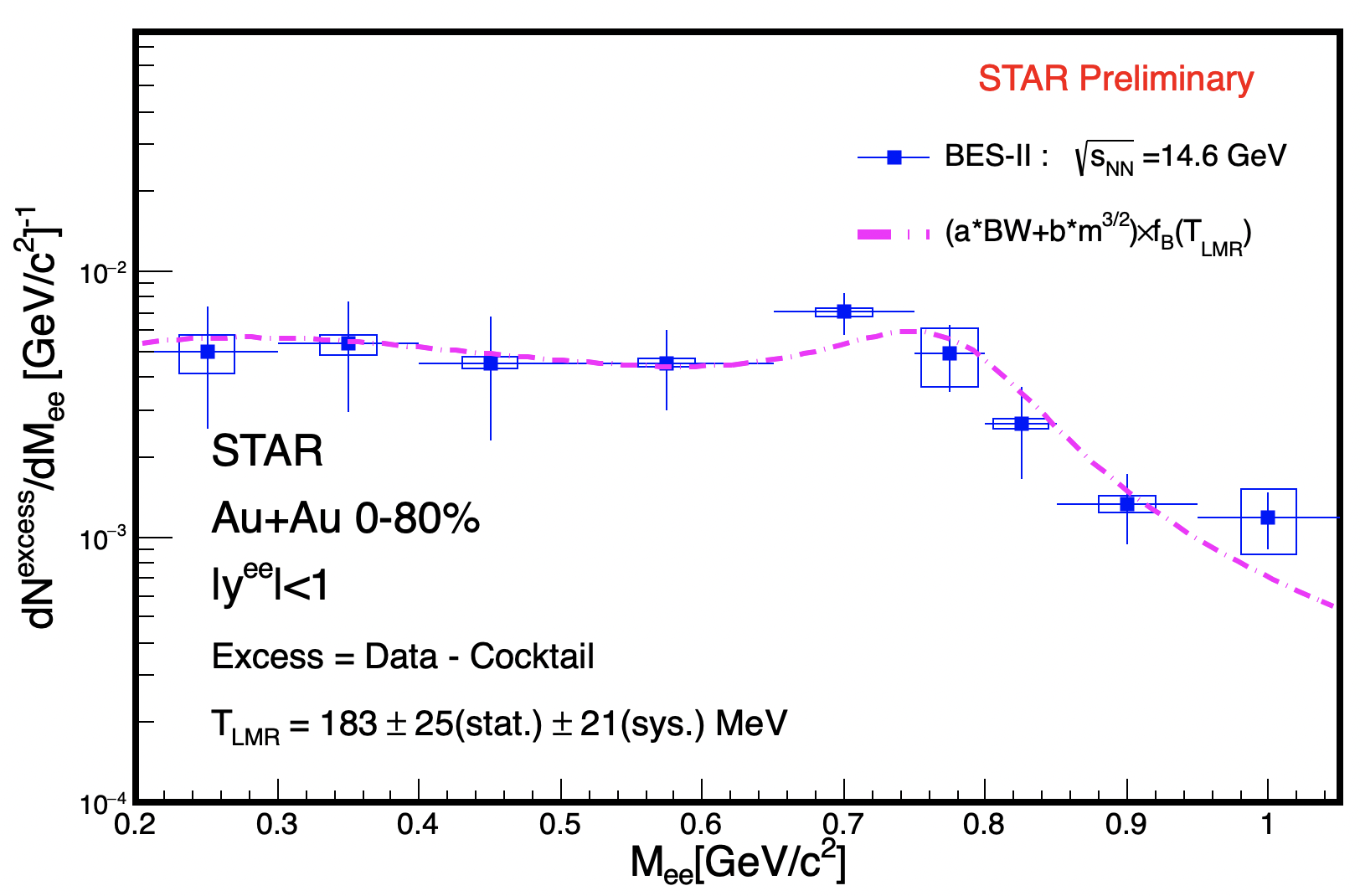}
\par\end{centering}

\protect\caption{Low invariant mass range (LMR) temperature ($M_{ll} <$ 1.1 GeV/$c^2$) fitting (dash line) via dielectron excess spectrum (points) at $\sqrt{s_{\text{NN}}}=$ 19.6 GeV (left) and $\sqrt{s_{\text{NN}}}=$ 14.6 GeV (right).}

\label{fig_bes2tem}
\end{figure}

A function that incorporates the in-medium resonance structure with the QGP radiation contribution is employed to fit the LMR spectrum: $(a\cdot BW+b\cdot M_{ee}^{3/2})*e^{-M_{ee}/T}$ \cite{Rapp:2014hha,STAR:2024review,STAR:2023tem}. BW represents Breit-Wigner function: $\frac{MM_{0}\Gamma}{(M_{0}^{2}-M^{2})^{2} + M_{0}^{2}\Gamma^{2}}$, where $M_{0}$ and $\Gamma_{0}$ are the pole mass and width of $\rho$ meson, while $M$ is the dielectron invariant mass. Fig. \ref{fig_bes2tem} presents the dielectron excess yields in LMR at $\sqrt{s_{\text{NN}}}=$ 19.6 and 14.6 GeV, fitted using equation above to extract effective temperature primarily from hadronic phase. The extracted temperatures are $168 \pm 13 ~\text{(stat.)} \pm 15 ~\text{(syst.)}$~MeV and $183 \pm 25 ~\text{(stat.)} \pm 21 ~\text{(syst.)}$~MeV for the Au+Au collisions at 19.6 and 14.6~GeV, respectively. 
\begin{figure}
\begin{centering}
\includegraphics[width=0.53\textwidth]{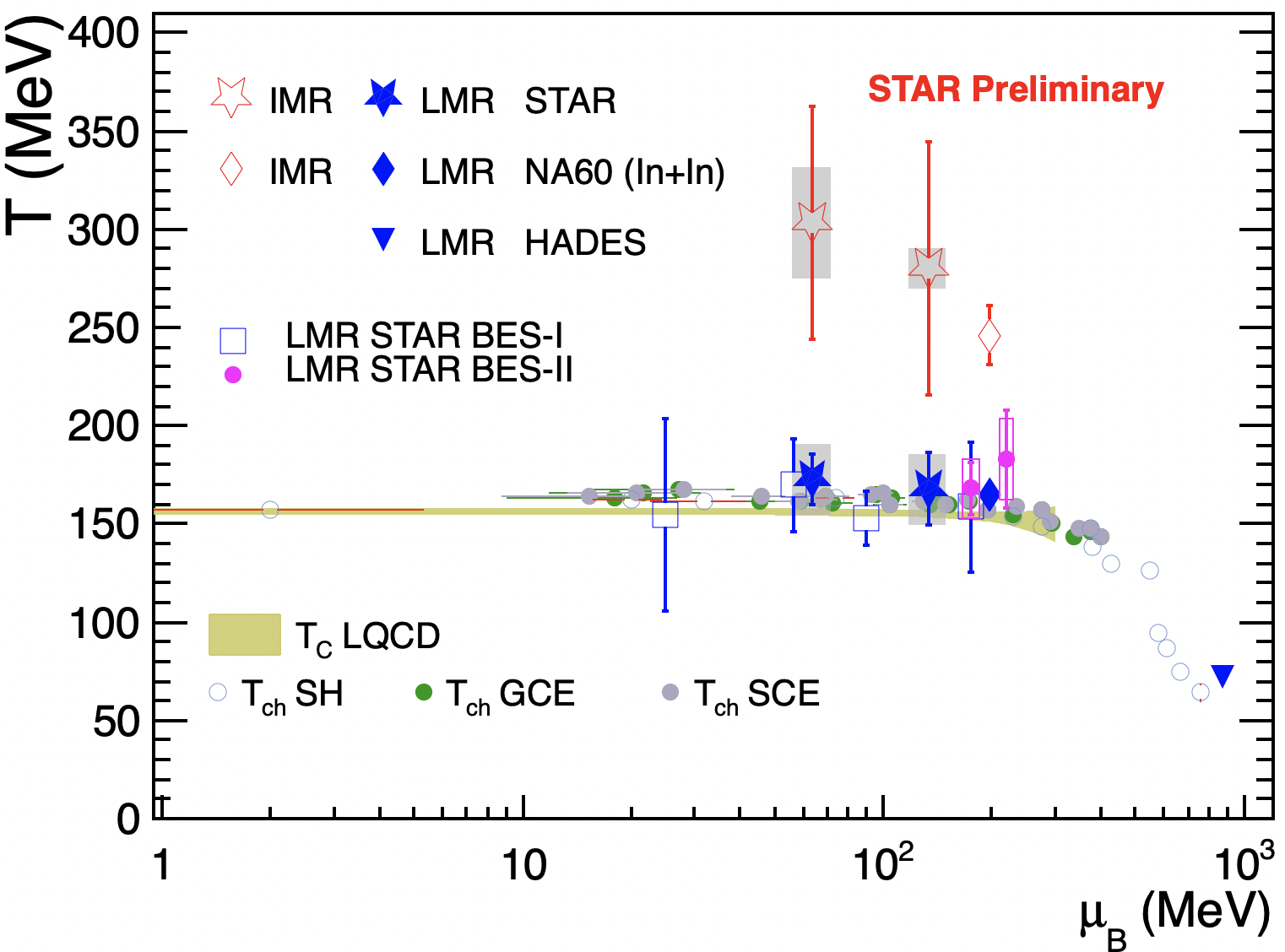}
\par\end{centering}

\protect\caption{\textbf{Temperatures vs. baryon chemical potential.} Temperatures extracted from LMR from STAR BES-II data (purple points) are  compared to the temperatures extracted from STAR 27 and 54 GeV data (stars) \cite{STAR:2023tem}, NA60 data (diamonds) \cite{NA60:2008dcb}, HADES data (inverted triangle) \cite{HADES:2019auv} and STAR BES-I data (blue open square). Chemical freeze-out temperatures extracted from the statistical thermal models (SH, GCE, SCE) are shown as open and filled circles \cite{STAR:2017sal, Andronic:2017pug}. The QCD critical temperature $T_{\rm{C}}$ at finite $\mu_{B}$ predicted by LQCD calculation is shown as a yellow band \cite{HotQCD:2018pds}.}

\label{fig_temsum}
\end{figure}

The summary of measured temperatures vs. baryon chemical potential is depicted in Fig. \ref{fig_temsum}. Temperatures from LMR are situated near pseudo-critical temperature, indicating thermal radiation from hadronic gas is predominantly generated in the vicinity of the phase transition.

\section{Summary}

In summary, the dielectron invariant mass spectra obtained from Au+Au collisions at $\sqrt{s_{\text{NN}}}=$ 7.7, 9.2, 11.5, 14.6, and 19.6 GeV are discussed. STAR's new observations of the integrated excess yield hint at a downward trend with decreasing collision energy, underscoring the medium conditions' impact on the EM spectral function and emphasizing the necessity of considering baryon chemical potential dependence in its modeling. In addition, the first temperature measurement at BES-II energies are presented, providing a unique access to the thermodynamic properties at the late stage near the phase transition to hadronic matter.


\end{document}